\RequirePackage{fixltx2e}
\documentclass[aps,prb,reprint,floatfix,superscriptaddress,longbibliography]{revtex4-1}
\usepackage{latexsym}
\usepackage{amsmath}
\usepackage{amsfonts}
\usepackage{amssymb}
\usepackage{amsthm}
\usepackage{physics}
\usepackage{enumitem}

\usepackage{graphicx}
\usepackage{dcolumn}
\usepackage{bm}
\usepackage{algorithm}
\usepackage{algpseudocode}
\usepackage{xcolor}
\AtBeginDocument{\usepackage{booktabs}} 

\DeclareMathOperator*{\argmax}{arg\,max}
\DeclareMathOperator*{\argmin}{arg\,min}

\newcommand{\eq}[1]{Eq.~(\ref{#1})} %
\def\be{\begin{equation}} %
\def\ee{\end{equation}} %
\def\bea{\begin{eqnarray}} %
\def\eea{\end{eqnarray}} %

\begin{document}

\author{Robert A. Lang}
\affiliation{Department of Physical and Environmental Sciences,
  University of Toronto Scarborough, Toronto, Ontario, M1C 1A4,
  Canada}
\affiliation{Chemical Physics Theory Group, Department of Chemistry,
  University of Toronto, Toronto, Ontario, M5S 3H6, Canada}
  
\author{Ilya G. Ryabinkin}
 \affiliation{OTI Lumionics Inc.,
  100 College St. \#351, Toronto, Ontario\, M5G 1L5, Canada}

\author{Artur F. Izmaylov}
\affiliation{Department of Physical and Environmental Sciences,
  University of Toronto Scarborough, Toronto, Ontario, M1C 1A4,
  Canada}
\affiliation{Chemical Physics Theory Group, Department of Chemistry,
  University of Toronto, Toronto, Ontario, M5S 3H6, Canada}
  \email{artur.izmaylov@utoronto.ca}

\title{Unitary transformation of the electronic Hamiltonian with an exact quadratic truncation of the Baker-Campbell-Hausdorff expansion} 

\begin{abstract}
Application of current and near-term quantum hardware to the electronic structure problem is highly limited by qubit counts, coherence times, and gate fidelities. 
To address these restrictions within the variational quantum eigensolver (VQE) framework, many recent contributions have suggested dressing the electronic Hamiltonian 
to include a part of electron correlation, leaving the rest to be accounted by VQE state preparation. We present a new dressing scheme that combines 
preservation of the Hamiltonian hermiticity and an exact quadratic truncation of the Baker-Campbell-Hausdorff expansion. The new transformation is 
constructed as the exponent of an involutory linear combination (ILC) of anti-commuting  Pauli products. It incorporates important strong correlation effects 
in the dressed Hamiltonian and can be viewed as a classical preprocessing step alleviating the resource requirements of the subsequent VQE application. 
{The assessment of the new computational scheme for electronic structure of the LiH, H$_2$O, and N$_2$ molecules shows significant increase in efficiency compared to conventional qubit coupled cluster dressings.}

\end{abstract}

\maketitle

\section{Introduction}
One of the desirable applications of near-term universal gate quantum computation is solving the electronic 
structure problem. Currently, the most feasible route to this task is through the variational quantum eigensolver (VQE), \cite{Peruzzo2014, Wecker2015} which is a hybrid technique involving an iterative minimization of the electronic energy expectation value 
\begin{align} \label{VQE_expectation}
E = \min_{\boldsymbol{\tau}} \bra{\bar 0} \hat U^\dagger (\boldsymbol{\tau}) \hat H \hat U (\boldsymbol{\tau})\ket{\bar 0},
\end{align}
 involving quantum and classical computers. First, a classical computer suggests a trial unitary transformation 
 $\hat U (\boldsymbol{\tau})$ that is encoded on a quantum computer as a circuit operating on 
the initial state of $n_q$ qubits $\ket{\bar 0} \equiv \ket{0}^{\otimes n_q}$. 
 Repeated executions of this circuit are performed to accumulate measurement statistics for parts of the qubit-space Hamiltonian $\hat H$, which is iso-spectral to the second-quantized electronic Hamiltonian. The total electronic energy estimate is provided to the classical optimizer that generates 
 a next guess for the variational parameters $\boldsymbol{\tau}$. 
 These cycles minimize the energy expectation value and serve to approach the true electronic ground state energy. The VQE procedure has been experimentally demonstrated for small molecular systems such as He-H$^{+}$, \cite{Peruzzo2014} LiH, \cite{Kandala:2017/nature/242, Hempel2018} BeH$_2$, \cite{Kandala:2017/nature/242} and H$_2$O. \cite{Ryabinkin2019a, Nam2020} 
  
There are several conceptual challenges in the VQE scheme that prevent its experimental application to chemical systems beyond small molecules. First, the qubit-space electronic Hamiltonian cannot be measured entirely and requires separate {repeated sampling} of its parts. The number of parts that can be measured simultaneously varies depending on the measurement technique, but it still scales at least as $n_q$ for small molecules. \cite{Izmaylov2019revising,Yen2019b,Izmaylov2019unitary,Verteletskyi2019,Zhao2019,crawford2019efficient, huggins2019efficient,CSA2020} 
Second, choosing an efficient unitary operator form and search for the global minimum of variational parameters 
$\boldsymbol{\tau}$ are exponentially hard problems that are left to heuristic processes on the classical computer. \cite{McClean2018, Cerezo2020} 
The generators of $n_q$-qubit unitary operations up to a global phase correspond to an exponentially large Lie algebra $\mathfrak{su}(2^{n_q})$, 
whose elements consist of all $n_q$-qubit Pauli products $\hat P$,
\begin{align} \label{pauli_prod}
\hat P = \bigotimes_{j=1}^{n_q} \hat \sigma_j,
\end{align}
where $\hat \sigma_j$'s are the Pauli operators  $\{\hat x, \hat y, \hat z \}$ or the $2 \times 2$ identity $\hat 1$ acting on the $j^{\rm th}$
qubit. The $\mathfrak{su}(2^{n_q})$ algebra contains $4^{n_q}-1$ basis generators, therefore one needs efficient strategies to make an optimal operator choice of
\bea
\hat U(\boldsymbol{\tau}) = \prod_{k} e^{-i\tau_k \hat P_k}
\eea
compatible with hardware limited by number of reliably performed quantum circuit operations. Note that quantum computers 
cannot perform $\exp(-i \tau_k \hat P_k)$ directly as elementary circuit operations (gates) 
but there are compiling schemes that present such operations as sequences of universal gates. \cite{Seeley2012}

Several forms for $\hat U$ have been suggested recently: 1) based on the fermionic unitary forms restricted by the 
orbital excitation level and then transformed to the qubit space,\cite{Romero2018, Peruzzo2014,Lee:2019/jctc/311,Sokolov2019,Evangelista:2019/arXiv/1910.10130}  
for example, unitary coupled cluster singles and doubles (UCCSD)\cite{Romero2018, Peruzzo2014,Sokolov2019,Mizukami2019,Hempel2018}; 
2) hardware based approaches, where the unitary transformation is obtained as a sequence of universal 
gates\cite{Kandala:2017/nature/242,Gard:2019vd,PhysRevApplied.11.044092};
and 3) constructive qubit-space techniques, where $\hat P_k$'s are selected based on the sensitivity of energy to variations of their amplitudes. \cite{Ryabinkin2018,Grimsley2019,qAdapt, Ryabinkin2019b}

{
In light of the limited qubit counts and circuit depths offered by contemporary quantum computers, many recent contributions have proposed to similarity-transform (\textit{dress}) the electronic Hamiltonian with the goal of alleviating the quantum resource requirements of the following VQE treatment in obtaining accurate energies. One strain of such works involves the explicitly correlated approach of trans-correlated (TC) Hamiltonians, where singularities arising from electron coalescence are suppressed by a dressing of the Hamiltonian, hence accelerating the basis-set convergence and thereby reducing the basis set error of fixed qubit count VQE calculations. A Slater-type geminal-based unitary trans-correlator has been recently employed towards this purpose, \cite{Motta2020} albeit, due to non-truncation of the TC Hamiltonian's Baker-Campbell-Hausdorff (BCH) expansion, approximate dressing was required consistent with the usual procedure of the canonical TC F-12 approach. \cite{Takeshi2012} Relatedly, a non-unitary Jastrow factor-based trans-correlator has been recently proposed for use in the context of quantum simulation, \cite{McArdle2020} with the TC Hamiltonian's BCH expansion truncating at second order. However, the non-unitary transformation leading to a non-hermitian TC Hamiltonian introduces its own difficulties for VQE application, namely invalidating the Rayleigh-Ritz variational principle. The non-variational nature of this TC Hamiltonian was circumvented by use of Hamiltonian ansatz-based quantum imaginary time evolution, \cite{HansatzQITE} requiring resource-demanding circuit execution of the time evolution operator.} 

{
In an effort to reduce the qubit requirements of larger basis set calculations employing the VQE, the double unitary coupled cluster (DUCC) formalism\cite{DUCCFormalism} was employed to dress the Hamiltonian with UCCSD excitations/de-excitations connecting the predefined active-space orbitals to the virtual-space, thereby incorporating dynamical correlation contributed by the virtual orbitals in the transformed Hamiltonian. \cite{Metcalf2020} The remaining part of the UCCSD operator, defined strictly within the active space, is then implemented using a quantum circuit, hence only requiring qubit counts adequate to describe the active space orbitals. However, approximate dressing was required due to the well-known non-truncation of the UCC-dressed Hamiltonian.\cite{Kutzelnigg1991}} 

{Alternatively, one could consider dressing the Hamiltonian with the traditional coupled cluster ansatz, whose BCH expansion is known to truncate at the quartic order commutators.\cite{Helgaker} Unfortunately, one ends up with the predicament of a non-hermitian operator which can not be applied variationally in a straightforward manner.
To mitigate circuit depths in the VQE, an iterative Hamiltonian dressing technique employing the qubit coupled cluster (QCC) ansatz was recently proposed,\cite{Ryabinkin2019b} where the Hamiltonian is dressed with an optimized product of singly exponentiated $\hat P_k$'s obtained through an energy lowering heuristic. Interestingly, the QCC dressing is both unitary and has a finite BCH expansion, facilitating variational treatment and exact dressing. Unfortunately, the growth of the dressed Hamiltonian still scales exponentially with the size of the QCC transformation.}

{
In this paper, we propose a dressing of the electronic Hamiltonian with a unitary transformation that results in only a quadratic multiplicative growth factor in the size of the dressed Hamiltonian. The unitary transformation is constructed by taking the exponent of an involutory linear combination (ILC) of anti-commuting  Pauli products. These Pauli products are obtained through an extended QCC operator screening procedure. Thus, the unitary transformation, which we refer to as the QCC-ILC transformation herein, incorporates the effects of the most energetically important electronic configurations (as identified by the QCC screening) into the dressed Hamiltonian. 
The QCC-ILC dressed Hamiltonian is then used as input to a subsequent VQE calculation with significantly reduced depth requirements to achieve fixed accuracy 
in the output energy expectation.
}

The rest of the paper is structured as follows. In Sec. \ref{QCC_theory} we review the standard QCC methodology and its  operator screening protocol. The general properties and conditions of the QCC-ILC unitary transformation are introduced in Sec. \ref{QCC-ILC_intro} and \ref{QCC-ILC_dress}. The explicit flow of the QCC-ILC algorithm, {which can be interpreted as a classical VQE pre-processing step}, is provided in Sec. \ref{sec:detailed_procedure}. In Sec.~\ref{sec:benchmarking}, capabilities of the QCC-ILC transformation are demonstrated on the dissociation potential energy curves (PECs) of LiH and H$_2$O obtained using VQE simulation of fixed quantum resources, {along with numerical assessment of the operator growth incurred by the QCC-ILC transformation applied to a N$_2$ electronic Hamiltonian}. We provide discussion along with concluding remarks and directions for future studies in Sec.~\ref{conclusion}.

\section{Theory}
\subsection{Qubit coupled cluster (QCC)} \label{QCC_theory}
We start with briefly reviewing some relevant details of the QCC approach.
The QCC ansatz \cite{Ryabinkin2018} is constructed as a product of entangling multi-qubit rotations acting on an unentangled reference state. The reference state is taken to be the qubit-mean field (QMF) wavefunction, defined as a product of single-qubit coherent states,
\begin{align}
\ket{\boldsymbol{\Omega}} & = \bigotimes_{i=1}^{n_q} \ket{\Omega_i}, \\
\ket{\Omega_i} & = \cos(\frac{\theta_i}{2}) \ket{0} + e^{i \phi_i} \sin(\frac{\theta_i}{2}) \ket{1},
\end{align} 
where $\boldsymbol{\Omega} \equiv \{ \theta_i, \phi_i \}_{i=1}^{n_q}$ 
are $2 n_q$ Bloch angles. The QCC wavefunction ansatz is 
\begin{align} 
\ket{\text{QCC}} & = \hat U_{\text{QCC}} \ket{\boldsymbol{\Omega}},  \label{QCC_wavefunction} \\ 
\hat U_{\text{QCC}} & = \prod_{i=1}^{N} e^{-i \tau_i \hat T_i/2},  \label{QCC_unitary}
\end{align}
where $\hat T_i$'s are Pauli products [\eq{pauli_prod}] containing Pauli operators for at least two qubits and are referred to herein as \textit{entanglers},  $\tau_i$'s are corresponding variational amplitudes, and $N$ is the number of entanglers 
included in the ansatz. The QCC energy is obtained by VQE optimization of the form
\begin{align}
E = \min_{\boldsymbol{\Omega}, \boldsymbol{\tau}} \bra{\boldsymbol{\Omega}} \hat U_{\text{QCC}}^\dagger (\boldsymbol{\tau}) \hat H \hat U_{\text{QCC}} (\boldsymbol{\tau}) \ket{\boldsymbol{\Omega}},
\end{align}
where $\boldsymbol{\tau} \equiv \{ \tau_i  \}_{i=1}^{N}$ is a set of variational amplitudes corresponding to $\{\hat T_i \}_{i=1}^{N}$ included in $\hat U_{\text{QCC}}$. The optimized QCC energy depends on the particular $\{ \hat T_i \}_{i=1}^{N}$ and their ordering in $\hat U_{\text{QCC}}$. Individual $\hat T_i$'s are selected and ordered based on the absolute value of their energy gradients, defined as 
\begin{align}
\frac{\partial E}{\partial \tau_i} \Big|_{\boldsymbol{\tau} = 0} & = - \frac{i}{2} \bra{\boldsymbol{\Omega}_{\text{min}}} [\hat H, \hat T_i] \ket{\boldsymbol{\Omega}_{\text{min}}}, \label{energy_gradient_expr}
\end{align}
where $\ket{\boldsymbol{\Omega}_{\text{min}}}$ denotes the lowest-energy QMF state, $\ket{\boldsymbol{\Omega}_{\text{min}}}  = \argmin_{\boldsymbol{\Omega}} \bra{\boldsymbol{\Omega}} \hat H \ket{\boldsymbol{\Omega}}$. We found that under some mild conditions on QMF, there is no need to screen the exponential number of entanglers ($O(4^{n_q})$), and one can construct directly the set of entanglers with large energy gradient magnitudes by a polynomially scaling classical algorithm. \cite{Ryabinkin2019b}

The condition on QMF that allows one to obtain the polynomial construction algorithm  
is that $\ket{\boldsymbol{\Omega}_\text{min}}$ has all qubits aligned along the $z$ quantization axis, i.e. $\forall i \> \theta_i \in \{0, \pi\}$. This corresponds (up to global phase) to $\ket{\boldsymbol{\Omega}_\text{min}} \in \{ \ket{0}, \ket{1} \}^{\otimes n_q}$, where $\{ \ket{0}, \ket{1} \}^{\otimes n_q}$ denotes the set of $n_q$-qubit computational basis states. By any fermion-to-qubit mapping {of the canonical orbitals,} 
$z$-collinear $\ket{\boldsymbol{\Omega}_{\text{min}}}$ corresponds to the qubit-space representation of { the Hartree-Fock state}. In cases where $\ket{\boldsymbol{\Omega}_{\text{min}}}$ is not a computational basis state, one can instead use {the nearest} computational basis state $\ket{\boldsymbol{\Phi}}$ for gradient evaluation \eq{energy_gradient_expr} which has the largest overlap with $\ket{\boldsymbol{\Omega}_\text{min}}$,
\begin{align} \label{purified_MF}
\ket{\boldsymbol{\Phi}} = \argmax_{\ket{\boldsymbol{\Phi}} \in \{ \ket{0}, \ket{1} \}^{\otimes n_q}} \left| \left \langle \boldsymbol{\Phi} | \boldsymbol{\Omega}_{\text{min}}  \right \rangle  \right|^2.
\end{align}
We refer to the set of all entanglers with nonzero absolute energy gradients evaluated on $\hat H$ with $\ket{\boldsymbol{\Phi}}$ as the \textit{direct interaction set} (DIS), denoted herein as $\mathcal{D} \left( \hat H,\boldsymbol{\Phi} \right)$. The DIS may be partitioned into $n_p$ disjoint subsets, $\mathcal{D} = \bigcup_{k=1}^{n_p} \mathcal{G}_k$, where each subset $\mathcal{G}_k$ contains $O(2^{n_q-1})$ entanglers of identical gradient magnitude. \cite{Ryabinkin2019b} The number of partitions $n_p$ for the qubit Hamiltonian scales linearly in the number of terms it contains in the fermionic representation, and hence efficient ranking of all entanglers in the DIS is accomplished by performing a gradient calculation only for a single representative entangler for each partition, thereby ranking the complete DIS of cardinality $O(2^{n_q-1}n_p)$ with merely $n_p$ gradient computations. 

After selecting $\hat T_i$'s on a classical computer 
and optimizing their amplitudes on a quantum or classical computer, one has a freedom to keep 
the obtained $\hat U_{\rm QCC}$ as a circuit or use it to transform the Hamiltonian. The latter 
approach is employed in the iterative QCC (iQCC) method, and leads to the exponential growth of the number of terms in the Hamiltonian with the number of 
$\hat T_i$'s in $\hat U_{\text{QCC}}$ but reduces the circuit depth requirements. Note that dressing $\hat H$ with each 
$\hat T_i$ multiplies the number of terms in $\hat H$ by only a constant factor 
because the BCH expansion of $\tilde H \equiv \exp(i \tau_i \hat T_i/2) \hat H \exp(-i \tau_i \hat T_i/2)$ is finite due to the involutory  property of entanglers $\hat T_i$, $\hat T_i^2 = \hat 1$, and
{
\begin{align} \label{QCC_dressing}
\tilde H = & \hat H + \sin{\tau_i} \left( - \frac{i}{2} [\hat H, \hat T_i]  \right) \nonumber \\ & + \frac{1}{2} \left( 1 - \cos{\tau_i} \right) \left( \hat T_i \hat H \hat T_i - \hat H  \right).
\end{align}
By recursive application of Eq.~(\ref{QCC_dressing}), it is clear that $\hat U_{\text{QCC}} \hat H \hat U_{\text{QCC}}^\dagger$ produces $3^{N}$ distinct terms for $\hat U_{\text{QCC}}$ containing $N$ entanglers. However, it was shown in Ref. \citenum{Ryabinkin2019b} that such a dressing results in a Hamiltonian $O(1.5^N)$ times larger than $\hat H$ in an average sense due to merging and cancellation of terms once numerical amplitudes are substituted. 
}

\subsection{Involutory linear combinations of entanglers} \label{QCC-ILC_intro}
{Motivated by the unitarity and finite BCH expansion of the QCC dressing, along with its success in lowering circuit depth requirements,\cite{Ryabinkin2019b} we aim to reduce the number of terms introduced in the Hamiltonian when dressed by a unitary similar to $\hat U_{\rm QCC}$ by constructing linear combinations 
of entanglers that are involutory.} 
Using the involutory property of entanglers, 
one can construct an \textit{involutory linear combination} (ILC) from set of entanglers $\mathcal{A} = \{\hat T_1, \hat T_2, \hdots  \}$,
\begin{align} \label{involutory_sum}
\left( \sum_{\hat T_i \in \mathcal{A}} \alpha_i \hat T_i \right)^2 = \hat 1,
\end{align}
under conditions of normalization,
\begin{align}
\sum_i \alpha_i^2 = 1, \label{normalization_cond}
\end{align}
and that all $\hat T_i \in \mathcal{A}$ are mutually anti-commutative,
\begin{align} \label{anticom_cond}
\{\hat T_i, \hat T_j \} = 0 \ \forall \ \hat T_i, \hat T_j \in \mathcal{A}.
\end{align}
With the two conditions satisfied, \eq{involutory_sum} allows for the encoding of a unitary linear expansion of identity and $|\mathcal{A}|$ mutually anti-commuting  entanglers as the exponent
\begin{align}
\hat U_{\text{ILC}} & = \exp(- i \tau \sum_{\hat T_i \in \mathcal{A}} \alpha_i \hat T_i)  \label{QCI_encoding_fund} \\
& = \cos(\tau) \hat 1 - i \sin(\tau)  \sum_{\hat T_i \in \mathcal{A}} \alpha_i \hat T_i \label{QCC-ILC_linear}.
\end{align}
{
Satisfying Eq.~(\ref{anticom_cond}) such that the included $\hat T_i$'s are variationally significant can be accomplished using the freedom in choosing elements from the DIS. Namely, every DIS partition $\mathcal{G}_k$ consists of $2^{{n_q}-1}$ operators of identical gradient magnitude. Selecting mutually anti-commuting elements from different partitions, it is possible to construct the desired ILC. However ILCs may not be made arbitrarily large, Sankar and van den Berg provided a rigorous upper bound of $ 2 n_q + 1 $ for the cardinality of sets of mutually anti-commuting $n_q$-qubit Pauli words. \cite{Paulis} Further, because generators from the DIS are additionally subjected to a condition that they must have non-zero energy gradients in Eq.~(\ref{energy_gradient_expr}) and hence include an odd number of $\hat y_i$ operators, the general estimate on maximal ILCs from the DIS is lowered to $2 n_q - 1$. Our strategy is to pick $N \leq 2 n_q - 1$ top-gradient DIS partitions and attempt to construct an anti-commutative set from their elements. In Appendix \ref{appendix:anticom_alg} we present a procedure for generating sets of anti-commutative elements from the desired DIS partitions, it requires not more than $O(n_q^6)$ binary arithmetic operations on a classical computer.
Since all entanglers within the same DIS partition $\mathcal{G}_k$ are mutually commutative, QCC-ILC mutually \textit{anti}-commutative entanglers are necessarily from different $\mathcal{G}_k$'s. In obtaining the optimized QCC-ILC transformation, {Eq.~(\ref{QCI_encoding_fund}) is optimized with respect to free parameter $\tau$ and constrained parameters $\{ \alpha \}_{i=1}^{|\mathcal{A}|}$ subject to Eq.~(\ref{normalization_cond})}. Due to the linearity of the ILC ansatz, \eq{QCC-ILC_linear}, one can efficiently obtain the optimal parameters using the linear variational method on a classical computer (further details are in Appendix \ref{appendix:combination_est}). 
}

\subsection{QCC-ILC unitary dressing} \label{QCC-ILC_dress}

The linearity of \eq{QCC-ILC_linear} has the main advantage of encoding $N$ high-gradient entanglers, generating energetically important configurations as determined by the QCC screening, while its expansion contains $N+1$ terms as opposed to $O(2^N)$ for the standard QCC unitary \eq{QCC_unitary}. A transformation of the Hamiltonian with a unitary of form \eq{QCI_encoding_fund} gives
\begin{align} \label{transformed_hamiltonian}
\hat U_{\text{ILC}} \hat H \hat U_{\text{ILC}}^\dagger = & \cos^2 (\tau) \hat H - \frac{i}{2} \sin(2 \tau) \sum_{i=1}^N \alpha_i [\hat H, \hat T_i] \nonumber \\
& + \sin^2 (\tau) \sum_{i=1}^{N} \sum_{j=1}^{N} \alpha_i \alpha_j \hat T_i \hat H \hat T_j.
\end{align}
{At worst, each commutator $[\hat H, \hat T_i]$ contains the same number of terms as $\hat H$, and therefore the summation over $N$ commutators in \eq{transformed_hamiltonian} contribute a growth factor of $N$ at most.} Rewriting the double summation in \eq{transformed_hamiltonian} as
\begin{align} \label{quadratic_sum}
\sum_{i=1}^{N} \sum_{j=1}^{N} \alpha_i \alpha_j \hat T_i \hat H \hat T_j = & \sum_{i=j}^N \alpha_i^2 \hat T_i \hat H \hat T_i \nonumber \\
& + \sum_{i>j}^N \alpha_i \alpha_j \left( \hat T_i \hat H \hat T_j + \hat T_j \hat H \hat T_i  \right),
\end{align}
it is clear that the diagonal summation does not introduce any new terms in $\hat H$. This is a consequence of the commutativity/anti-commutativity of all Pauli terms and the involutory property of $\hat T_i$'s. 
The $\hat T_i \hat H \hat T_j$ and $\hat T_j \hat H \hat T_i$ can be seen to generate the same Pauli terms up to sign differences, 
and hence the last summation in Eq.~(\ref{quadratic_sum}) introduces an $N(N-1)/2$ growth factor at worst. 
The total worst-case multiplicative growth factor $G_{\text{worst}}$ for dressing with an $N$-entangler QCC-ILC unitary is 
\begin{align}
G_{\text{worst}}(N) = 1 + N + N(N-1)/2 = \frac{1}{2} \left(N^2 + N + 2 \right). \label{growth_factor_worst}
\end{align}
{
The average case scaling may be approximately obtained under the assumption that all Pauli products in the Hamiltonian, along with the entanglers entering $\hat U_{\text{ILC}}$, are uniformly sampled from the set of $4^{n_q}$ Pauli products [Eq.~(\ref{pauli_prod})]. An $n_q$-qubit Pauli product commutes with half of all $4^{n_q}$ Pauli products while anti-commuting  with the remaining half, leading to the summation over $[\hat H, \hat T_i]$ in Eq.~(\ref{transformed_hamiltonian}) contributing an $N/2$ growth factor in expectation. By similar reasoning, each $\hat T_i \hat H \hat T_j + \hat T_j \hat H \hat T_i$ in the second summation in Eq.~(\ref{quadratic_sum}) contributes a growth factor of $1/2$ on average if $[\hat T_i, \hat P_k] = 0$ and $[\hat T_j, \hat P_k] = 0$ for a given $\hat P_k$ in $\hat H$ with independent probabilities $1/2$, leading to the summation Eq.~(\ref{quadratic_sum}) contributing a growth factor of $N(N-1)/4$. The total average-case multiplicative growth factor $G_{\text{avg}}$ is then
\begin{align} \label{growth_factor_avg}
G_{\text{avg}}(N) = \frac{1}{4} \left(N^2 + N + 4 \right) = \frac{G_{\text{worst}}(N)+1}{2}.
\end{align}
While the Pauli products in $\hat H$ along with selected $\hat T_i$ admit more structure than being uniformly random in actuality, we numerically find that Eq.~(\ref{growth_factor_avg}) yields excellent Hamiltonian growth estimates in Sec.  \ref{sec:benchmarking}.
}

\subsection{The QCC-ILC procedure} \label{sec:detailed_procedure}
{
The procedure for obtaining the optimized QCC-ILC transformation, given an input fermion-to-qubit mapped electronic Hamiltonian, and number of entanglers $N$ to enter $\hat U_{\text{ILC}}$, is outlined herein.
\begin{enumerate}[wide, labelwidth=!, labelindent=0pt]
\item Obtain the DIS $\mathcal{D}(\hat H, \boldsymbol{\Phi})$ by the QCC operator ranking procedure.\cite{Ryabinkin2019b}
\item Generate mutually anti-commuting  subset $\mathcal{A} \subset \mathcal{D}(\hat H, \boldsymbol{\Phi})$ with $|\mathcal{A}| = N \leq 2 n_q - 1$ using techniques outlined in Appendix \ref{appendix:anticom_alg}.
\item Obtain the optimal $N$ free parameters for $\hat U_{\text{ILC}} = \exp(-i \tau \sum_{\hat T_i \in \mathcal{A}} \alpha_i \hat T_i)$ for qubit mean-field reference $\ket{\boldsymbol{\Phi}}$ with optional qubit mean-field relaxation using techniques outlined in Appendix \ref{appendix:combination_est}.
\item Dress the Hamiltonian as $\hat H' = \hat U_{\text{ILC}}^\dagger (\tau, \boldsymbol{\alpha}) \hat H \hat U_{\text{ILC}} (\tau, \boldsymbol{\alpha})$ [Eq.~(\ref{transformed_hamiltonian})] using the fixed ILC parameters found in the previous step.
\end{enumerate}
The resulting Hamiltonian $\hat H'$ can then be used for an ensuing VQE treatment. Optionally, the QCC-ILC dressing procedure can be repeated, taking the resulting $\hat H'$ of Step $4$ as an updated input to Step $1$. While $\hat H'$ is only $O(N^2)$ times larger than $\hat H$ in the number of its Pauli terms, repeating the QCC-ILC dressing procedure $d$ times will of course result in a final transformed Hamiltonian $O(N^{2d})$ times larger than the initial $\hat H$.
}
{Prior to running the QCC-ILC procedure, one can get an accurate estimate of the number of Pauli terms in $\hat H'$ by multiplying the number of terms in the initial Hamiltonian by $G_{\text{avg}}$ [Eq.~(\ref{growth_factor_avg})].} If one wishes to employ exponential procedure of $d$ repeated transformations, the growth factor of the final Hamiltonian can be approximately taken as $G_{\text{avg}}^d$. Alternatively, $d$ may be determined \textit{a posteriori} by convergence of the optimized QCC-ILC energies [see Eq.~(\ref{QCC_ILC_energy})] or gradient magnitudes in $\mathcal{D}$. For instance, if the QCC-ILC energy difference between consecutive iterations falls below a given threshold, or the maximum gradient magnitude for $\mathcal{D}$ in Step 1 is below a threshold, then exit the dressing procedure. 

{Due to the QCC-ILC transformation's linear parameterization, it does not depend on the order of $\hat T_i$'s. However, if one is to re-calculate the DIS along a potential energy surface (PES), it is possible that the top $N$ partitions change. Thus, if one naively recalculates the DIS at each nuclear geometry and constructs the QCC-ILC transformation from the top $N$ partitions, it is possible to introduce significant kinks in the resulting PES, an inherent property of constructive or selective ans\"{a}tze. To avoid the possibility of energetic kinks arising from ansatz adaptation, one can perform DIS calculation and subsequent ILC construction at a single geometry, and use the resulting ansatz for optimization over the entire PES. The point in which the DIS is computed can be taken to be any nuclear geometry where electron correlation is expected to be significant. Since the gradient magnitudes are perturbative measures of correlation energy, one can calculate the DIS magnitudes along the PES, and perform ILC construction using the highest $N$ partitions where gradient magnitudes are maximized.}

\section{Numerical results}  \label{sec:benchmarking}
{
Within this section, we numerically elucidate various properties of the introduced QCC-ILC transformation for molecules in variously sized active spaces. The capability of the QCC-ILC transformation employed as a classical overhead to subsequent VQE application is illustrated for the potential energy curves (PECs) of LiH dissociation and symmetric bond stretch of H$_2$O in the STO-$3$G and $6$-$31$G bases respectively in Sec. \ref{PECs_subsection}. In Sec. \ref{N2_growth_sec}, the increase in the number of Hamiltonian Pauli terms resulting from random QCC-ILC transformations is benchmarked against random QCC transformations of similar specification for N$_2$ in the cc-pVDZ basis. In Sec. \ref{H2O_growth_comp_sec}, a Hamiltonian growth comparison is conducted using optimally selected generators for the QCC-ILC and standard QCC ans\"{a}tzes, using their optimized amplitudes for dressing, along with a benchmark of their corresponding energy estimates, for H$_2$O in the $6$-$31$G(d) basis. When Hamiltonian Pauli term counts are reported, only terms with coefficient magnitudes exceeding $10^{-8}$ are accounted for.
}

\subsection{Increasing accuracy of fixed-depth VQE} \label{PECs_subsection}
To assess the capabilities of the QCC-ILC transformation in increasing the electronic energy estimates for VQE optimizations using limited quantum resources, we perform the QCC-ILC procedure as an overhead to standard QCC VQE application. We use QCC-ILC transformed Hamiltonian in obtaining simulated QCC VQE energies along the PECs for dissociation of LiH and symmetric dissociation of H$_2$O in the STO-$3$G and $6$-$31$G basis sets respectively. Active space fermionic Hamiltonians for the LiH and H$_2$O systems were generated and transformed to their qubit space representations by parity and Bravyi-Kitaev transformations respectively. An active space of $2$ electrons in $3$ molecular orbitals, CAS($2$e, $3$o) was employed for LiH, and CAS($4$e, $4$o) for H$_2$O. Following qubit tapering procedures, \cite{tapering_qubits} the resulting LiH and H$_2$O Hamiltonians act on $4$- and $6$-qubit spaces respectively. All details on the system Hamiltonians can be seen in Table 1 of Ref \citenum{Ryabinkin2019b}. Reported energy errors are with respect to the CASCI (complete active space CI) energies obtained through exact diagonalization of the active-space Hamiltonians. 

For proof of concept, we utilize repeated QCC-ILC dressing. While repeating the QCC-ILC procedure is bound to result in an exponential blowup in the number of Pauli terms of the dressed Hamiltonian, we demonstrate that it is possible to increasingly improve VQE energy estimates limited by fixed-depth quantum circuits. We see that generally the first QCC-ILC dressing will capture the majority of electronic correlation, and hence the polynomially scaling $d=1$ procedure is of the greatest utility. We denote the sequence of $d$ QCC-ILC dressings of $N$ entanglers each followed by subsequent VQE simulation using QCC ansatz of $M$ entanglers as iQCC-ILC($d$, $N$, $M$), due to its resemblance of the iterative QCC (iQCC) procedure. \cite{Ryabinkin2019b} 

%
The QCC ansatz is constructed from selecting entanglers from the $M$ highest gradient magnitude partitions of the DIS. For the QCC ans\"{a}tze constructed for the final QCC-ILC dressed Hamiltonians, such entanglers are generated by re-evaluation of the DIS, and constructing a representative for the top $M$ partitions by placing $\hat x$ instances on every qubit index with a non-zero $x$ component in the $x$ vector characterizing the partition, except for a single $\hat y$ instance at the first index with non-zero $x$ component to satisfy the odd $\hat y$-parity. In cases where $M$ is greater than the number of DIS partitions $n_p$, the remaining $M - n_p$ entanglers are generated by permutations of $\hat x$ with $\hat y$ in the parent $n_p$ entanglers. 

Once $M$ entanglers have been generated for inclusion in the QCC ansatz, an operator ordering strategy must be prescribed, due to non-commutativity in the product Eq.~(\ref{QCC_unitary}). We employ ordering by ascending gradient magnitude (i.e. highest gradient magnitude entangler acts directly on the QMF reference). To alleviate the problem of discontinuous PECs, the entire iQCC-ILC procedure is performed only once at a single geometry where initial DIS gradient magnitudes are sufficiently high, whereas all other geometries do not feature re-selection of QCC-ILC entanglers or final QCC entanglers, only re-optimization of amplitudes and sequential dressing are performed.

\subsubsection{LiH dissociation}
\begin{figure*}
\centering
\hspace{-1.3cm}
\includegraphics[width=0.8\textwidth]{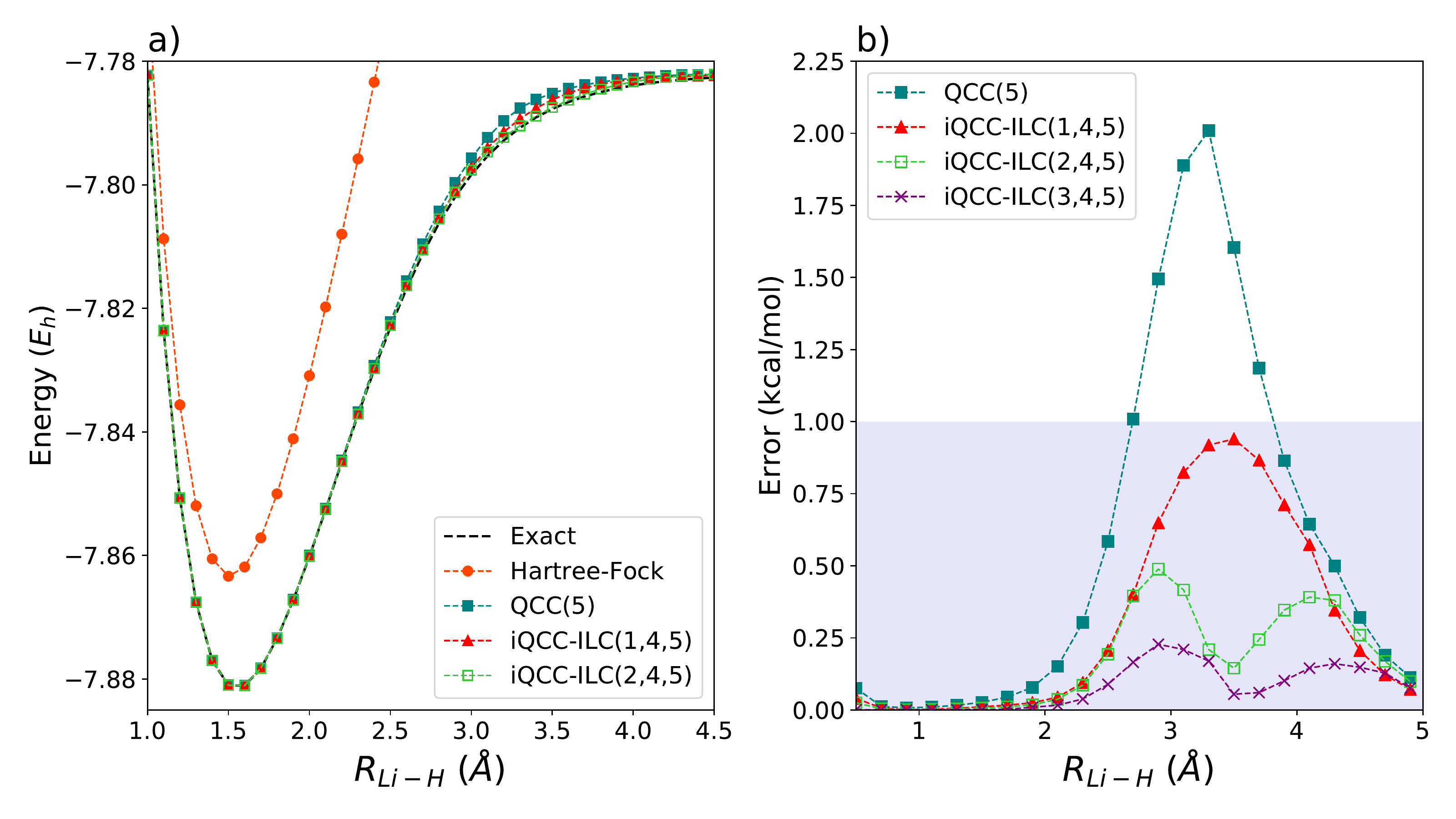} 
\caption{a) PECs for the ground state CAS($2$e, $2$o) LiH dissociation in different methods. QCC(5) ansatz uses the bare qubit Hamiltonian and 5 entanglers ($\hat x_1 \hat y_3$, $\hat x_1 \hat y_3 \hat x_4$, $\hat x_1 \hat x_2 \hat y_3 x_4$, $\hat x_1 \hat y_2 \hat x_3 $, and $\hat x_1 \hat y_2 \hat x_3 \hat x_4 $). b) Ground state energy errors along the LiH dissociation for different methods. The shaded area denotes errors within chemical accuracy.}
\label{LiH_qcc_ilc} 
\end{figure*}

We apply the iQCC-ILC procedure in producing LiH dissociation PECs in STO$3$-G basis seen in Figure \ref{LiH_qcc_ilc}a. All iQCC-ILC PECs are generated with dressings of $4$-entangler ILC unitaries (with intermediate relaxation of QMF Bloch angles permitted, i.e. optimization of $4$ free ILC amplitudes and $8$ Bloch angles) and final $5$-entangler QCC VQE optimization ($5$ entangler amplitudes and $8$ Bloch angles). 

Selection of QCC-ILC and QCC entanglers are performed at geometry $R = 3.0~$\AA, where correlation is sufficiently high, $E_{\text{corr}} \approx 55 \text{ kcal}/\text{mol}$. The entanglers selected during the iQCC-ILC procedure for this geometry are then applied for the complete PEC. This way, energy discontinuities are alleviated by preserving the obtained DIS along the entire PEC. While the QCC ansatz with $5$ entanglers and no dressings is seen to be qualitatively sufficient, 
an error of up to $2 \text{ kcal}/\text{mol}$ is observed in Figure \ref{LiH_qcc_ilc}b. 
A single dressing with a QCC-ILC unitary of $4$ entanglers with subsequent $5$-entangler QCC VQE is seen to produce 
energy estimates within the chemical accuracy for the entire PEC.

Further dressing is seen to improve the energy estimates for the majority of the curve. 
{Sometimes, iQCC-ILC($d$, $N$, $M$) gives higher energy than iQCC-ILC($d-1$, $N$, $M$), 
for example, the iQCC-ILC($2,4,5$) error is marginally ($0.02-0.05$ kcal/mol) above the iQCC-ILC($1, 4, 5$) 
error at extended bond length $R > 4.2~\text{\AA}$. This is a result of more efficient energy lowering by 
5 entanglers selected based on a single-dressed Hamiltonian than by their counterparts obtained for a 
double-dressed Hamiltonian. These results do not violate the variational character of the method since 
they are obtained for two differently dressed Hamiltonians.}
Generally, the maximum energy error is seen to lower for the majority of the curve with 
sequential dressings with entanglers selected at a single geometry.

With initial LiH qubit Hamiltonians of $100$ terms, a single dressing produces Hamiltonians consisting of $135$ terms. 
A second and further dressings yield a transformed Hamiltonian of $136$ terms which appears to be the limiting value. 
This is the result of conserving the time-reversal symmetry of the original Hamiltonian by 
the QCC-ILC transformations. Time-reversal symmetry requires even number of $\hat y$ operators in 
individual Pauli products and reality of their coefficients.  
The number of possible even $\hat y$-parity $n_q$-qubit Pauli terms is
\begin{align}
\sum_{m=0}^{\left \lfloor \frac{n_q}{2} \right \rfloor} \binom{n_q}{2m} 3^{n_q - 2m},  \label{even_y_parity}
\end{align}
which gives $136$ for the $n_q=4$ case. 

\subsubsection{H$_2$O symmetric dissociation}

\begin{figure*}
\centering
\hspace{-1.3cm}
\includegraphics[width=0.8\textwidth]{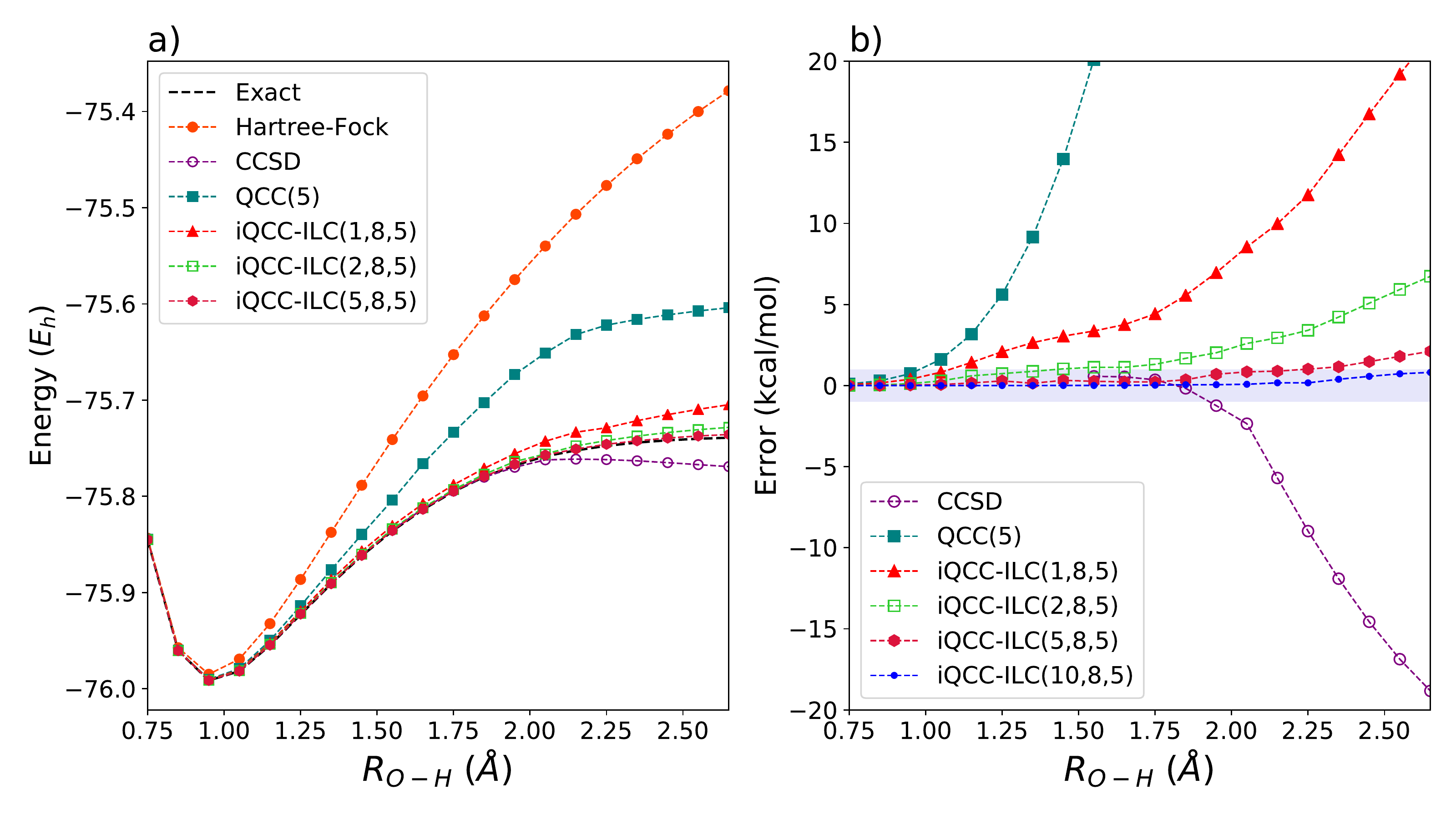} 

\caption{a) PECs for the ground state of CAS($4$e, $4$o) H$_2$O along the symmetric bond stretch in different methods. QCC(5) employs the
spin-penalized Hamiltonian and 5 entanglers ($\hat x_2 \hat y_3 \hat x_5 \hat x_6$, $\hat x_2 \hat y_3 \hat x_4 \hat x_5$, $\hat x_1 \hat y_3$, $\hat x_1 \hat y_2 \hat x_5 \hat x_6$, and $\hat x_1 \hat y_2 \hat x_4 \hat x_5$).  
b) Ground state energy errors along the symmetric bond stretch for different methods. The shaded area denotes errors within chemical accuracy.}
\label{H2O_qcc_ilc}
\end{figure*}

We perform benchmarking of the iQCC-ILC procedure for the symmetric O-H bond dissociation of CAS($4$e, $4$o) H$_2$O in $6$-$31$G basis. The simultaneous dissociation of both O-H bonds along the symmetric stretching of H$_2$O is a well-known case of 
strong correlation and hence presents as an interesting test for the iQCC-ILC ansatz within the strongly correlated regime. 
We employ iQCC-ILC($d$, $8$, $5$) schemes and consider varying numbers of transformations up to $d=10$ 
in Figure \ref{H2O_qcc_ilc}, to demonstrate the systematic improvability 
of energy estimations with fixed-size quantum circuits via the QCC-ILC transformation. 

To improve convergence to the singlet ground state, all calculations here were performed starting with a spin-penalized Hamiltonian,
\begin{align}
\hat H_s = \hat H + \frac{\mu}{2} \hat S^2,
\end{align}
where $\mu$ is a positive constant ($\mu = 0.5 E_h$). This extra term serves as a penalty on non-singlet spin contamination during optimization. 
Since $\hat S^2$ is a two-body operator, $\hat H_s$ and $\hat H$ contain the same number of terms.
All entangler selection was performed at $R = 2.35~$\AA{} where correlation energy is significant ($\sim 185$ kcal/mol).

The initial Hamiltonian $\hat H_s$ for water contains $165$ terms. The transformed Hamiltonian dressed with a single QCC-ILC unitary of $8$ entanglers obtained in the iQCC-ILC($1$, $8$, $5$) procedure contains $995$ terms, demonstrating a $\sim 6$ times growth factor. This is quite low in relation to the average growth factor $G_{\text{avg}} = 19$ predicted by \eq{growth_factor_worst} due to near-saturation of maximum Hamiltonian terms. Further dressings increase the number of Hamiltonian terms to $1055$, at which point the number of terms saturates. A saturation point of $1055$ terms is lower than predicted by \eq{even_y_parity}. However, the dressed Hamiltonians are seen to have not only even $\hat y$-parity but also even $\hat x$-parity, for which there are $1056$ such Pauli products in the $n_q=6$ qubit Pauli product algebra. The singly-dressed iQCC-ILC($1$, $8$, $5$) procedure is seen to {significantly improve upon the PEC accuracy for the symmetric dissociation with respect to the standard QCC procedure of $5$ entanglers} for the spin-penalized Hamiltonian, evident in Figure \ref{H2O_qcc_ilc}a. However, the iQCC-ILC($1$, $8$, $5$) procedure yields an error of $> 20 \text{ kcal}/\text{mol }$ for the end of the considered reaction coordinate. Systematically improved energy estimates are obtained with further dressings. The iQCC-ILC procedure with $10$ transformations demonstrates microHartree accuracy for the majority of the PEC, with chemical accuracy achieved for the entire considered reaction coordinate. The iQCC-ILC($2$, $8$, $5$) procedure is already seen to energetically outperform the coupled cluster with single and double excitations (CCSD) result at extended geometry (Figure \ref{H2O_qcc_ilc}b), the latter produces significant non-variational behaviour.

\begin{table}[htb!]
\caption{Comparison of variously configured iQCC-ILC, iQCC, and QCC procedures for CAS($4$e, $4$o) H$_2$O at $R = 2.35 \text{\AA}$ (strongly correlated regime), using the spin-penalized Hamiltonian. Notation iQCC($d$, $N$, $M$) denotes an iQCC procedure of $d$ sequential dressings with optimized $N$-entangler QCC ans\"{a}tze with final $M$-entangler QCC optimization. State $\Psi$ refers to the final normalized trial state resulting from the respective procedure.}
\begin{tabular*}{\columnwidth}{@{\extracolsep{\fill}} l c c c}
\toprule 
Method                &    \multicolumn{1}{p{1.5cm}}{\centering Terms in \\ final $\hat H_s$}   & $|\left \langle \Psi | \text{FCI} \right \rangle |$  & \multicolumn{1}{p{1.5cm}}{\centering $E_{\Psi} - E_{\text{FCI}}$  \\ (kcal/mol)}                        \\ \hline
HF                            & $165$						& $0.6330$												  &   $185.09$                            \\
QCC($10$)				      & $165$						& $0.9570$												   &  $13.36$						\\
iQCC-ILC($1$, $8$, $5$)    	  & $995$            			& $0.9526$                                                  &	  $13.59$                      \\ 
iQCC($1$, $8$, $5$)            & $923$                       &  $0.9562$                                                 &   $15.54$                        \\
iQCC-ILC($1$, $8$, $10$)    	  & $995$            			& $0.9784$                                                  &	 $6.10$                     \\ 
iQCC($1$, $8$, $10$)            & $923$                       &  $0.9896$                                                 &   $5.20$                   \\
\hline
\end{tabular*} \label{strongcorr_data}
\end{table}
{Various iQCC-ILC procedures characterized by a single dressing step ($d=1$) applied to water at $R = 2.35$\AA{} are compared along with standard QCC and iQCC procedures in Table \ref{strongcorr_data}. The presence of strong correlation at this geometry is evident by the low overlap of the Hartree-Fock state with the FCI ground state. To be precise, an iQCC procedure, a single dressing with an optimized $N$-entangler QCC ansatz with final $M$-entangler QCC optimization is denoted as iQCC($1$, $N$, $M$). {Unlike the QCC-ILC transformation, the iterative optimization of the QCC unitaries must be done on a quantum circuit via VQE due to exponential scaling of the QCC energy functional on a classical computer. The iQCC($d$, $N$, $M$) procedure requires nonlinear optimization of the QCC ansatz at each $(d+1)^{\rm{th}}$ step, with each step in principle involving many function calls, each of which needing many measurements to reach a tolerable precision error. In contrast, the iQCC-ILC($d$, $N$, $M$) instead only requires a nonlinear optimization of the final QCC ansatz via VQE, while all $d$ optimal ILC dressings are sequentially obtained via linear variational method on a classical computer.} It is apparent for this system that the single-dressing iQCC and iQCC-ILC procedures with the same $N$ and $M$ values give comparable Hamiltonian growth, energy errors, and FCI overlaps. The comparable sizes of the dressed Hamiltonians for the QCC-ILC dressing and standard QCC dressing is deemed to be a symptom of small qubit count, i.e. the Hamiltonian term saturation point is already in vicinity of the singly dressed Hamiltonian sizes for both transformations. We apply both dressings in the following subsections to numerically assess the Hamiltonian growth systematically for larger Hilbert spaces. Most notable from Table \ref{strongcorr_data} is that the $8$-entangler QCC-ILC transformation with subsequent application of $5$-entangler QCC optimization gives energy and FCI overlap very comparable to that of QCC with $10$ entanglers applied to the initial Hamiltonian. Hence, comparable solution quality is achieved with half the number of QCC entanglers, alleviating both circuit depth and optimization difficulty in the VQE calculation. Applying the $10$-entangler QCC VQE ansatz after the QCC-ILC transformation shows that significantly higher quality energy estimates and FCI overlaps can be achieved using fixed-size VQE ans\"{a}tze.
}

{
\subsection{Assessment of Hamiltonian growth} \label{N2_growth_sec}
Within this subsection we compare increase in the number of Pauli terms in $\hat H$ under dressing with standard QCC and QCC-ILC unitary transformations. 
To avoid confusion, here, we will refer to the latter as the ILC transformation hereafter. As illustrated in Table \ref{strongcorr_data}, the ILC- and QCC-dressed Hamiltonians have comparable numbers of Pauli terms for fixed $N$ when applied to the CAS($4$e, $4$o) active space water Hamiltonian in $6$-$31$G basis. The comparable scaling of the Hamiltonian growth is presumed to be a symptom of the low upper bound on number of terms which can enter the transformed Hamiltonian for $n_q = 6$. To numerically assess the Hamiltonian growth resulting from ILC unitary transformations compared with standard QCC transformations, we employ transformations on an N$_2$ electronic Hamiltonian in the cc-pVDZ basis in an active space of $6$ electrons in $6$ spatial orbitals, CAS($6$e,$6$o). Molecular spin orbitals are enumerated using the same-spin grouping, and the $12$ qubit Hamiltonian consisting of $247$ Pauli terms is obtained via parity transformation of the canonical fermionic creation/annihilation operators. 
\begin{figure}
\centering
\hspace*{-0.6cm} \includegraphics[width=0.53\textwidth]{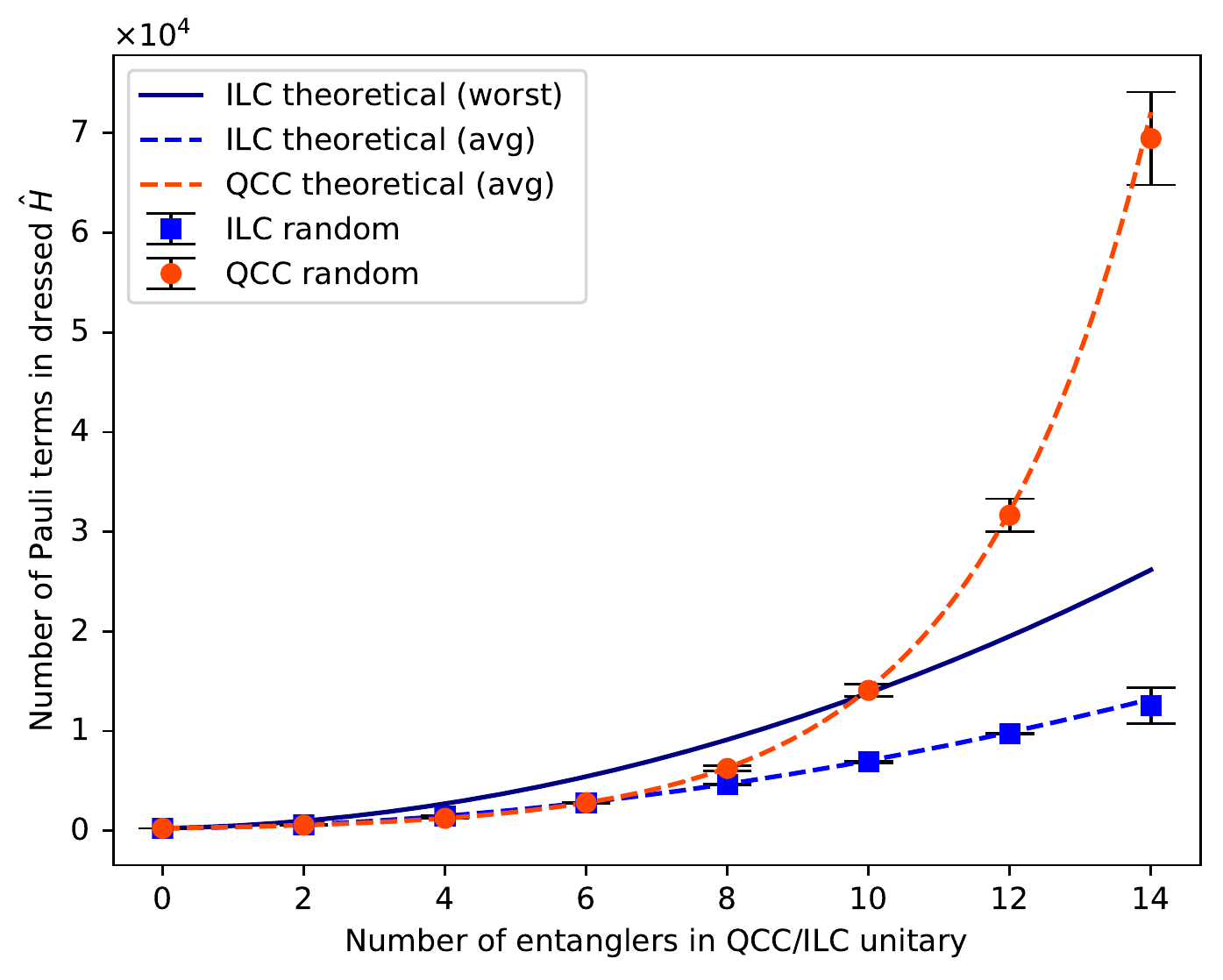} 
\caption{Comparison of theoretical and random case Pauli term growth of the CAS($6$e, $6$o) N$_2$ initial Hamiltonian ($\hat H_{\text{N}_{2}}$) in the cc-pVDZ basis after QCC($N$) and ILC($N$) unitary transformations. Points denote the average number of Pauli terms in the dressed Hamiltonians, and error bars denote standard deviations. The number of Pauli terms in $\hat H_{\text{N}_{2}}$ is $m=247$. The worst and average case theoretical estimates for the ILC transformation are given by $mG_{\text{worst}}$ and $mG_{\text{\text{avg}}}$ (see Eqs. (\ref{growth_factor_worst}) and (\ref{growth_factor_avg})) respectively. The average case theoretical QCC transformation size estimate is $m(3/2)^N$.\cite{Ryabinkin2019b}}
\label{growth_comparison_n2} 
\end{figure}   
The typical growth incurred by the dressings is numerically assessed by performing uniformly random QCC and ILC transformations  on the initial Hamiltonian with varying number of entanglers ($N$) entering the QCC/ILC unitaries, and plotted in Figure \ref{growth_comparison_n2}. A random QCC transformation of $N$ entanglers is obtained by uniformly choosing $N$ Pauli words of odd $y$-parity, and uniformly setting their associated amplitudes in the interval $[0, 2\pi)$. A random ILC transformation of $N$ anti-commuting entanglers is obtained from the following procedure: 1) $N$ uniformly random bit-strings of length $n_q$ are generated and used to represent a set of $N$ DIS flip indices (describing placement of $\hat x_i$'s and $\hat y_i$'s), 2) the procedure of Appendix \ref{appendix:anticom_alg} is then used to find a set of $N$ anti-commuting entanglers from the flip-indices by solving Eq.~(\ref{gauss_elim_problem}), 3) finally the global amplitude of the ILC unitary is then selected uniformly at random, $\tau \in [0, 2\pi)$, and the remaining parameters $\{\alpha\}_{i=1}^N$ are taken to be the components of a random $N$-dimensional unit vector. The averages and standard deviations in Figure \ref{growth_comparison_n2} are obtained from $10$ trials of the uniformly random dressing procedures for each value of $N$. 
The average number of Pauli terms in the QCC- and ILC-transformed Hamiltonians are comparable for $N \leq 6$, suggesting that the benefit of reduced Hamiltonian growth of ILC dressings over QCC dressings manifests only for larger dressing instances. The exponential and polynomial growth of QCC and ILC dressing respectively becomes evident at $N \approx 10$.  The standard deviations for both the QCC and ILC dressings are remarkably low relative to the averages with only $10$ trials performed, suggesting the number of Pauli terms in the dressed Hamiltonians is predominantly determined by the number of entanglers entering the unitary, rather than which specific entanglers are chosen and the values of their associated amplitudes. Therefore, the average-case multiplicative growth factor Eq.~(\ref{growth_factor_avg}) can be used to obtain estimates for the size of QCC-ILC transformed Hamiltonians to high accuracy. 
} 

{
\subsection{Performance comparison to QCC dressing} \label{H2O_growth_comp_sec}
To assess how the ILC energy compares to QCC energies in regimes where the ILC dressed Hamiltonians become sufficiently smaller than QCC dressed Hamiltonians, we perform benchmarking using a full valence active space H$_2$O electronic Hamiltonian in the $6$-$31$G(d) basis. From the total $19$ atomic orbitals, only the $12$ $sp$-type HF molecular orbitals were included to form the active space of CAS($8$e, $12$o), i.e. freezing the oxygen $1s$ core and the remaining $d$-character virtual orbitals. The H$-$O$-$H angle was set to $107.6^\circ$, and a symmetric O$-$H bond length of $R = 1.5$ \AA{} was used.  The resulting $24$ qubit Hamiltonian consisted of $8921$ Pauli terms after the JW mapping.}

{
We compare optimized ILC energies and transformed Hamiltonian growth factors with those of various optimal QCC ans\"{a}tze in Figure \ref{qcc_comparison_h2o_full_valence}. An optimal QCC ansatz of $N$ entanglers, QCC($N$), is conventionally formed by selecting one element from each of the $N$ DIS partitions of largest gradient magnitude, and ordered by ascending gradient magnitude when read from left to right in Eq.(\ref{QCC_unitary}). Since each partition contains $O(2^{n_q-1})$ entanglers, there is great freedom in choosing specific entanglers to enter the QCC unitary. While any choice of Pauli words from the top $N$ DIS partitions is considered to be optimal by the first-order gradient heuristic, the final optimized energy and transformed Hamiltonian size will be a function of the specific Pauli word selections. Thus, in order to assess how the ILC transformation of $N$ anti-commuting Pauli words, ILC($N$), compares to analogous QCC($N$) transformations, we construct 500 samples of top-gradient QCC($N$) ans\"{a}tze by randomly selecting a DIS element from each of the top $N$ partitions. All ans\"{a}tze use the gradient magnitude ordering. Each resulting QCC energy functional was optimized using the limited-memory Broyden-Fletcher-Goldfarb-Shanno (L-BFGS) algorithm. \cite{BFGS} For $N=6$, one can find a set of anti-commuting Pauli words from the top $6$ DIS partitions to generate the first-order optimal ILC($6$) transformation. However, the top $8$ and $10$ DIS partitions here do not permit mutually anti-commuting solutions, and hence one must use a set of non-optimal partitions to construct the anti-commuting set from. 

\begin{figure*}[htb!]
\centering
\hspace{-1.8cm}
\includegraphics[width=0.9\textwidth]{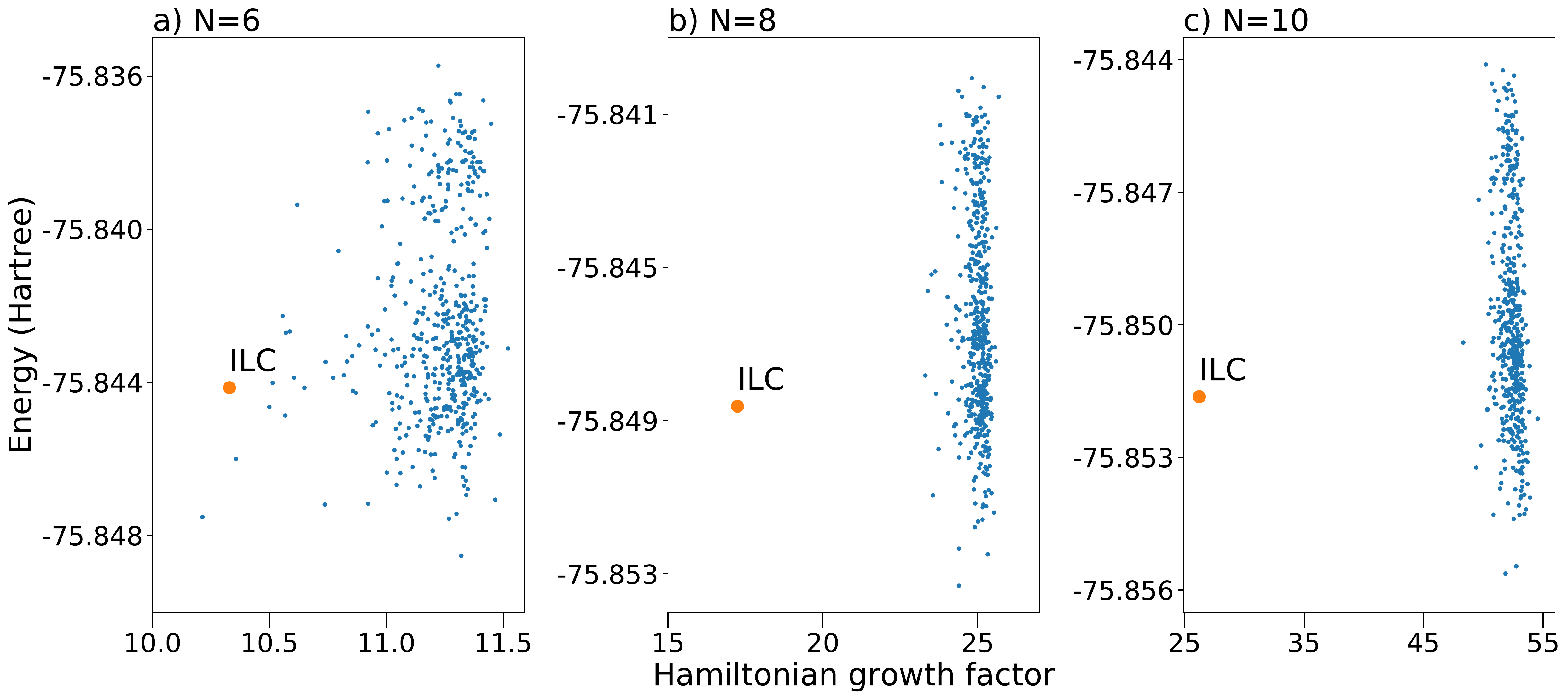}
\caption{Comparison of optimized energies and dressed Hamiltonian growth for the ILC($N$) ansatz and $500$ randomly chosen first-order optimal QCC($N$) ans\"{a}tze (blue points) applied to the CAS($8$e, $12$o) H$_2$O electronic Hamiltonian in the $6$-$31$G(d) basis at  $R = 1.5$ \AA. The Hamiltonian growth factors are taken as the ratio of the number of Pauli terms in the dressed Hamiltonian to the number of Pauli terms in the initial Hamiltonian ($8921$).}
\label{qcc_comparison_h2o_full_valence} 
\end{figure*} 

The average performance of the first-order optimal QCC($N$) ans\"{a}tze compared  with the ILC($N$) ansatz are displayed in Table \ref{average_qcc_comparison_h2o_full_valence}. Pauli term counts for the dressed Hamiltonians using the ILC($N$) transformation become drastically lower than those for the average of first-order optimal QCC($N$) dressing for increasing $N$, with nearly a factor of $2$ reduction in growth for $N=10$. By extrapolation one can expect furtherly drastic reductions for $N>10$. The tabulated relative energies indicate that using the ILC transformation can significantly reduce  Pauli term growth of the dressed Hamiltonian while robustly maintaining marginally lower energies when compared to the average over first-order optimal QCC unitary ans\"{a}tze of similar specification for large scale calculations.

\begin{table}
\caption{Comparison of the optimized energies and the number of terms in the dressed Hamiltonian for the ILC($N$) ansatz and the average of the first-order optimal QCC($N$) ans\"{a}tze obtained from $500$ samples (see Figure \ref{qcc_comparison_h2o_full_valence}). The energies for the ILC($N$) ansatz and the average of the QCC($N$) ans\"{a}tze are denoted $E_{\text{ILC(}N\text{)}}$ and $E_{\text{QCC(}N\text{)}}^{(avg)}$. The number of Pauli terms in the dressed Hamiltonians for the ILC($N$) ansatz and the average of the QCC($N$) ans\"{a}tze are denoted $m_{\text{ILC(}N\text{)}}$ and $m_{\text{QCC(}N\text{)}}^{(avg)}$ respectively.}
\begin{tabular*}{\columnwidth}{@{\extracolsep{\fill}} l c c c}
\toprule 
$N$ & \multicolumn{1}{p{1.5cm}}{\centering $m_{\text{QCC(}N\text{)}}^{(avg)}$  \\ $\times 10^{-3}$} & \multicolumn{1}{p{1.5cm}}{\centering $m_{\text{ILC(}N\text{)}}$ \\ $\times 10^{-3}$} & \multicolumn{1}{p{3cm}}{\centering $E_{\text{ILC(}N\text{)}} - E_{\text{QCC(}N\text{)}}^{(avg)}$ \\ (milliHartree)}  \\ \hline
$6$  & $100$ & $92$ & $-1.72$ \\ \hline
$8$  & $223$ & $154$ & $-2.34$ \\ \hline
$10$  & $467$ & $234$ & $-1.76$ \\ \hline
\hline
\end{tabular*} \label{average_qcc_comparison_h2o_full_valence}
\end{table}
}
{
While the implemented ILC($8$) and ILC($10$) unitaries are not first-order optimal, they still provide energies slightly lower than the first-order optimal QCC($8$) and QCC($10$) averages respectively. Since one cannot find a set of anti-commuting Pauli words from the top $8$ ($10$) DIS partitions here, a brute force search through the DIS partitions is performed until a set is identified which mutually anti-commuting Pauli words can be found from, i.e. for which a solution exists in the procedure outlined in Appendix \ref{appendix:anticom_alg}. The identified set of $8$ ($10$) partitions turn out to have very similar gradient magnitudes to the top $8$ ($10$). The similar gradient magnitudes between the set of top $N$ and the set of anti-commutative set-constructable $N$ partitions explains the comparable optimized energies up to a first-order consideration. One can expect this to be a common scenario in the viscinity of orbital quasi-degeneracies: there will be many DIS partitions of quasi-degenerate gradient magnitude, and hence one can find many sets of $N$ partitions that have gradient magnitudes similar to those of the top $N$ set of partitions.
}


\section{Concluding remarks} \label{conclusion}
We have introduced a new Hamiltonian dressing technique with the motivation of alleviating quantum resource requirements of the following VQE treatment. The QCC-ILC transformation is unitary, allowing for straightforward variational application. Further, the exact dressing results in only a quadratic growth factor in the number of Pauli terms in the fermion-to-qubit mapped electronic Hamiltonian by use of exponentiated ILCs of QCC entanglers. 
The linearity of the QCC-ILC unitary allows for efficient parameter optimizations on a 
classical computer, further relieving quantum resources of the scheme. One can thus view the QCC-ILC optimization procedure as an efficient classical pre-processing technique that alleviates subsequent VQE-style optimization. 

The QCC-ILC dressing is seen to aid significantly in yielding accurate energy 
estimates using fixed circuit depth QCC ans\"{a}tze particularly in the strongly correlated regime (e.g. symmetric bond stretch in H$_2$O). 
In this regime, the single-reference state $\ket{\boldsymbol{\Phi}}$ used in the undressed QCC entangler 
selection procedure has small overlap with the highly correlated ground state of interest. 
This leads to fast growth of the number of necessary entanglers in the QCC scheme and to unfavorable circuit depths.  
The optimized QCC-ILC transformation of the Hamiltonian has the important benefit of effectively 
using multi-reference state $\hat U_{\text{ILC}} \ket{\boldsymbol{\Phi}}$ for the subsequent QCC ranking  
by including $\hat U_{\text{ILC}}$ at the operator level rather than the state. The QCC-ILC transformation was seen to capture significant amounts of correlation energy, particularly in the strongly correlated regime, as seen for the simultaneous bond-dissociation of CAS($4$e, $4$o) H$_2$O in the $6$-$31$G basis.  {Systematic reduction in Hamiltonian Pauli term growth from the QCC-ILC transformation compared to analogous QCC dressings was demonstrated via random entangler selection and amplitude sampling for CAS($6$e, $6$o) N$_2$ in the cc-pVDZ basis, and first-order optimal entangler selection with optimized amplitudes for CAS($8$e, $12$o) H$_2$O in the $6$-$31$G(d) basis.}

To elucidate the QCC-ILC transformation's capability of incorporating strong correlation, it is useful to consider what the QCC-ILC unitary corresponds to in the fermionic representation. By any standard fermion-to-qubit mapping such as JW or BK, there is a correspondence between the Pauli products and the Majorana products. The Majorana operators in a given single-particle basis may be written
\begin{align}
\hat \gamma_{2i} & = \hat a_i^\dagger + \hat a_i  \label{Maj_even}\\
\hat \gamma_{2i+1} & = i \left(\hat a_i^\dagger - \hat a_i \right) \label{Maj_odd}.
\end{align}
It is obvious from Eqs.~(\ref{Maj_even}) and (\ref{Maj_odd}) that a product of $k$ Majorana operators acting on different spin-orbital indices may be written as a linear combination of $2^k$ fermionic strings of length $k$, which takes a configuration to a superposition of varying particle number ranging from the $-k$ to $+k$ subspaces. Note that rank $k/2$ fermionic excitations/de-excitations are included in this linear combination for even $k$. When the QCC entangler ranking quantity Eq.~(\ref{energy_gradient_expr}) is taken with respect to the Hartree-Fock state and the initial electronic Hamiltonian $\hat H$, then non-zero gradient $\hat T_i$ must be including fermionic excitations with rank no higher than $2$, since $\hat H$ is a two-body operator. Further, $\hat T_i$  including single excitations have zero gradient as a result of Brillouin's theorem. Hence $\hat T_i$ with non-zero gradient must include fermionic double excitations, i.e. $\hat T_i$ can always be written up to a phase as a quartic Majorana product,
\begin{align} \label{fermionic_entangler}
\hat T_i = \hat n \prod_{j=1}^4 \hat \gamma_{a_j},
\end{align}
with $a_u < a_v + 1$ for $u < v$, and where $\hat n$ may be defined as 
\begin{align}
\hat n = \prod_{l} (1 - 2 \hat a_l^\dagger \hat a_l) = \prod_{l} \hat \gamma_{2l} \hat \gamma_{2l+1},
\end{align}
with the product over $l$ being a subset of the spin orbital indices. Inclusion of $\hat n$ in Eq.~(\ref{fermionic_entangler}) is generally required, since, when acting on a single configuration, $\hat n$ produces a global phase $\{1, -1 \}$, for which the gradient magnitude is invariant to. Generally, Majorana products can break virtually all symmetries of $\hat H$ unless careful linear combinations of them are constructed. In fact, ILCs of mutually anti-commuting  Majorana products can not be adapted to satisfy electronic Hamiltonian symmetries.\cite{IzmaylovOrdering} The QCC-ILC unitary obtained for the electronic Hamiltonian can therefore be considered as a selected symmetry-broken coupled cluster doubles (CCD) unitary ansatz. The additional constraint of all Majorana products being mutually anti-commuting  admits a quadratic sized BCH expansion of the exactly dressed Hamiltonian. Variants of CCD theory have previously seen  success in accurately describing strongly correlated systems. \cite{Krylov1998, VanVoorhis2000, VanVoorhis2000q, Limacher2013, Stein2014} We attribute the ability of the QCC-ILC transformation to incorporate strong correlation in the Hamiltonian to this connection to CCD theories, along with its ability to break virtually all symmetries. One may find it desirable to restore symmetries after the QCC-ILC dressing. \cite{Qiu2018} While exact symmetries can be enforced by initially dressing a symmetry-projected Hamiltonian,\cite{Yen2019a} one may be able to obtain more sophisticated routes of symmetry-restoration in the form of a unitary operation. A symmetry-restoring unitary is inherently more challenging to obtain than a symmetry projector, since its functional form will depend on the symmetry-broken input state. We leave the search for such symmetry-restoring unitaries for a future contribution.

Furthermore, the introduced unitary transformation is strictly energetically equivalent to the linear variational form introduced in Ref. \citenum{RyabinkinENPT}, and therefore is 1) globally optimized with polynomially scaling efforts, 2) variational, and 3) size-consistent due to its exponential parameterization.
\footnote{Size-consistency of the QCC-ILC transformation may only be rigorously guaranteed if a single dressing is performed. Otherwise, one may potentially find erroneous couplings between non-interacting subsystems due to the product of non-commutative ILC unitaries.} 
The current bottleneck in the QCC-ILC procedure is obtaining the set of anti-commutative entanglers from the desired DIS partitions, requiring high polynomial time relative to system size ($O(n_q^6)$ in the worst-case). A further detriment is that a solution is not guaranteed for any set of DIS partitions, and hence in its current formulation, one may sometimes need to exhaustively search until an adequate set is found. We expect there to be more efficient means to finding anticommuting elements from the DIS, we intend to explore such possibilities in a future work.

\section*{Acknowledgements}
 A.F.I acknowledges financial support from the Google Quantum Research Program 
 and the Natural Sciences and Engineering Research Council of Canada (NSERC). 
 R.A.L acknowledges graduate student funding from NSERC.

\section*{Appendix}

\appendix

\section{Generating mutually anti-commuting  sets of entanglers} \label{appendix:anticom_alg}
{In the standard QCC entangler selection procedure, one obtains the DIS partitions $\mathcal{G}_k$ from which arbitrary entanglers can be chosen, typically taking one entangler from each of the $N$ highest gradient $\mathcal{G}_k$. For the ILC transformation, the constraint of mutual anti-commutativity poses additional difficulty in the entangler selection process. A brute force search through the considered $\mathcal{G}_k$'s is unfeasible since each $\mathcal{G}_k$ contains $2^{n_q-1}$ entanglers, likewise heuristic graph techniques\cite{Izmaylov2019unitary} are unfavorable due to the exponential number of vertices. Instead, we make use of a binary vector representation of Pauli words and formulate the problem of finding $N$ anti-commuting entanglers from the DIS partitions of interest as a system of linear equations.
}

{In the binary vector representation, any $n_q$-qubit Pauli word $\hat P_k$ may be represented by two binary vectors of dimension $n_q$, $\vec z^{\> (k)}$ and $\vec x^{\> (k)}$, where their $l^{\rm{th}}$ components specify the single-qubit Pauli operator acting on the $l^{\rm{th}}$ qubit in $\hat P_k$, \cite{tapering_qubits}
\begin{align}
(x^{(k)}_l , z^{(k)}_{l}) = \begin{cases} (0,1) \ \> \text{$l^{\rm th}$ qubit is $\hat z$} \\ (1,0) \ \> \text{$l^{\rm th}$ qubit is $\hat x$} \\ (1,1) \ \> \text{$l^{\rm th}$ qubit is $\hat y$}  \\ (0,0) \ \> \text{$l^{\rm th}$ qubit is identity}  \end{cases}.
\end{align}
Noting that all entanglers within the same $\mathcal{G}_k$ have identical $\vec x$ vectors, we can formulate the problem of finding mutually anti-commuting elements from partitions $\{ \mathcal{G}_k \}_{k=1}^N$ as follows: given $\{ \vec x^{\> (k)} \}_{k=1}^N$, the set of $\vec x$ vectors characterizing $\{ \mathcal{G}_k \}_{k=1}^N$ in a one-to-one manner, find the set of $\vec{z}$ vectors $\{ \vec z^{\> (k)} \}_{k=1}^N$ such that the set of entanglers $\{ \hat T_k \}_{k=1}^N$ represented by $\{(\vec x^{\> (k)}, \vec z^{\> (k)}) \}_{k=1}^N$ are 1) mutually anti-commutative and 2) possessing an odd number of $\hat y$ instances. The condition of mutual anti-commutativity produces $N(N-1)/2$ linear equations on the $N n_q$ free variables in $\{ \vec z^{\> (k)} \}_{k=1}^N$ of the form
\begin{align} \label{constraint_anticom}
A_{jk} \equiv \sum_{l=1}^{n_q} x^{(j)}_{l} z^{(k)}_{l} +  z^{(j)}_{l} x^{(k)}_{l} \mod 2 = 1, 
\end{align}
where $1 \leq j < k \leq N$. Note that Eq.~(\ref{constraint_anticom}) is equivalent to the $2n_q$-dimensional concatenated vectors $(\vec x^{\> (j)} \ \vec z^{\> (j)})$ and $(\vec x^{\> (k)} \ \vec z^{\> (k)})$ being non-orthogonal with respect to the symplectic inner product over the binary field, which implies corresponding $\hat P_j$ and $\hat P_k$ anti-commute. \cite{tapering_qubits} The condition that all $\{ \hat T_k \}_{k=1}^N$ have an odd number of $\hat y$ instances is easily enforced by demanding that each  $(\vec x^{\> (k)}, \vec z^{\> (k)})$ pair has an odd number of $(x^{(k)}_l , z^{(k)}_l)$ occurrences. As $N$ linear equations of the $\{ \vec z^{\> (k)} \}_{k=1}^N$ variables,
\begin{align}
O_k  \equiv \sum_{l=1}^{n_q} x^{(k)}_l z^{(k)}_l \mod 2 = 1,
\end{align}
where $1 \leq k \leq N$. Collectively, $\{ A_{jk} \}_{j < k}^N$ and $\{ O_k \}_{k=1}^N$ are $N(N+1)/2$ binary linear equations, for which finding the $N n_q$ free variables in $\{ \vec z^{\> k} \}_{k=1}^N$ which satisfy their system yields $\{ (\vec x^{\> (k)}, \vec z^{\> (k)}) \}_{k=1}^N \to \{ \hat T_k \}_{k=1}^N$ which are mutually anti-commuting elements of the corresponding input DIS partitions $\{ \mathcal{G}_k \}_{k=1}^N$.
}

{
The linear system of equations can be formulated in matrix notation, for which standard implementations of binary gaussian elimination can be readily employed for finding a solution, if one exists. For instance, let $\boldsymbol{z}$ denote the $(N n_q)$-dimensional concatenated $\vec {z}$ vectors, $\boldsymbol{z} = \left( \vec {z}^{\> (1)} \ \hdots \ \vec {z}^{\> (N)} \right)^T$, and let $\boldsymbol{1}$ denote the $[N(N+1)/2]$-dimensional vector with unit value in all components. Then the system of linear equations can be written
\begin{align} \label{gauss_elim_problem}
\boldsymbol{A} \boldsymbol{z} = \boldsymbol{1},
\end{align} 
where $\boldsymbol{A}$ is a binary matrix of $N n_q$ columns and $N(N+1)/2$ rows. Each of the $N(N-1)/2$ anti-commutative constraints $A_{jk}$'s are encoded in a row of $\boldsymbol{A}$. The row corresponding to constraint $A_{jk}$ has the form $( \boldsymbol{0} \hdots \vec x^{(k)} \hdots \boldsymbol{0} \hdots \vec x^{(j)} \hdots \boldsymbol{0})$ where $\boldsymbol{0}$ is the $n_q$-dimensional zero vector, $\vec x^{(k)}$ and $\vec x^{(j)}$ indices range from $[(j-1)n_q + 1]$ to $jn_q$ and $[(k-1)n_q + 1]$ to $kn_q$ respectively. The remaining $\hat y$ parity constraints are encoded in the remaining $N$ rows of $\boldsymbol{A}$, where row corresponding to constraint $O_k$ takes the form $( \boldsymbol{0} \hdots \vec x^{\> (k)} \hdots \boldsymbol{0} )$, with $\vec x^{\> (k)}$ from column indices $[(k-1)n_q + 1]$ to $kn_q$.
}

Finite field Gaussian elimination requires $O(m^3)$ binary arithmetic operations for $m$ variables, \cite{Mezard2009} hence obtaining solution to Eq.~(\ref{gauss_elim_problem}) can be done in $O(N^3 n_q^3)$ time. Since $N$ is bound by $O(n_q)$, the worst-case scaling of this procedure is $O(n_q^6)$. There exist instances of $\boldsymbol{A}$ for which $\boldsymbol{1}$ is not in its image, i.e. Eq.~(\ref{gauss_elim_problem}) has no solution. In cases where no solution exists, we employ greedy method of replacing the lowest gradient $\mathcal{G}_k$ considered with the next-lowest gradient $\mathcal{G}_k$ until a solution to the corresponding $\boldsymbol{A}$ can be found. If there exists no next-lowest gradient $\mathcal{G}_k$ in the greedy process, the lowest is simply removed, and an effective $N' = N - 1$ is used.

\section{Optimization of the QCC-ILC unitary} \label{appendix:combination_est}

The state generated by acting a QCC-ILC unitary, $\hat U_{\text{ILC}}$, on the reference computational basis state (Slater determinant) $\ket{\boldsymbol{\Phi}}$ may be written
\begin{align}
\hat U_{\text{ILC}} \ket{\boldsymbol{\Phi}} & = \exp(- i \tau \sum_{i=1}^N \alpha_i \hat T_i ) \ket{\boldsymbol{\Phi}} \\ 
  & \equiv c_1 \ket{\boldsymbol{\Phi}} - i \sum_{j=2}^{N+1} c_j \hat T_j \ket{\boldsymbol{\Phi}}, \label{CI_formulation}
\end{align}
where $c_1 = \cos(\tau)$ and $c_{j} = \sin(\tau) \alpha_{j-1}$. Letting $\ket{\Phi_1} \equiv \ket{\boldsymbol{\Phi}}$, and $\ket{\Phi_j} \equiv \hat T_j \ket{\boldsymbol{\Phi}}$ for $j > 1$, one has by construction,
\begin{align}
\left \langle \Phi_i | \Phi_j \right \rangle = & \delta_{ij},
\end{align}
since all $\hat T_j$ in the involutory combination are necessarily characterized by different flip indices (they originate from different DIS partitions and hence have different placements of $\hat x_i$ and $\hat y_i$ operators) and $\ket{\boldsymbol{\Phi}}$ is a computational basis state. Finding the optimal amplitudes and energy minimum can hence be done by generating Hamiltonian matrix $\boldsymbol{H}$ in the set of $N+1$ orthogonal basis functions $\{ \ket{\Phi_j} \}_{j=1}^{N+1}$,
\begin{align}
H_{ij} \equiv \bra{\Phi_i} \hat H \ket{\Phi_j},
\end{align}
from which the energy minimum and optimal parameters $\{c_j \}_{j=1}^{N+1}$ can be obtained via standard orthogonal eigen-problem formulation. The imaginary phase of the summed over configurations in \eq{CI_formulation} can be explicitly accounted for by modifying the Hamiltonian matrix by element-wise multiplication, $\bar{\boldsymbol{H}} = \boldsymbol{M} \circ \boldsymbol{H}$, where 
\begin{align}
\boldsymbol{M} = \begin{pmatrix} 1 & \boldsymbol{-i}_{1 \times N} \\ \boldsymbol{i}_{N \times 1} & \boldsymbol{1}_{N \times N} \end{pmatrix}.
\end{align}
The optimal coefficients $\{ c_i \}_{i=1}^{N+1}$ in \eq{CI_formulation} may then be obtained by solving for the ground state solution of
\begin{align} \label{eigenproblem}
\bar{\boldsymbol{H}} \boldsymbol{c} = E \boldsymbol{c},
\end{align}
where $E$ corresponds to the energy minimum for the QCC-ILC unitary,
\begin{align} \label{QCC_ILC_energy}
E = \min_{\tau, \boldsymbol{\alpha}} \bra{\boldsymbol{\Phi}} \hat U_{\text{ILC}}^\dagger (\tau, \boldsymbol{\alpha}) \hat H \hat U_{\text{ILC}} (\tau, \boldsymbol{\alpha}) \ket{\boldsymbol{\Phi}}.
\end{align}
Unique matrix elements of $\boldsymbol{H}$ may be efficiently obtained on a classical computer due to the polynomial scaling of evaluating QMF expectation values as trigonometric polynomial functions of the $2n_q$ Bloch/Euler angles. \cite{Genin2019} Once \eq{eigenproblem} has been solved, the QCC-ILC parameters $\{ \tau, \alpha_1 , \hdots \alpha_N \}$ may be extracted from ground eigenvector $\boldsymbol{c}$ by obtaining amplitude $\tau$ as
\begin{align}
\tau = \arccos(c_1),
\end{align}
which can then be used to obtain $\{ \alpha_i \}_{i=1}^N$,
\begin{align} \label{alphas_from_betas}
\alpha_{j-1} = \frac{c_{j}}{\sin{\tau}}.
\end{align}
From \eq{alphas_from_betas}, $\{ \alpha_i \}_{i=1}^N$ are undefined when $\tau = 0$ (modulo $\pi$), however, such a scenario does not happen in practice. This is a result of all entanglers being selected from the DIS, and hence have non-zero energy gradient evaluated at $\tau = 0$, ensuring the optimized $\tau$ amplitude will be non-zero.

The procedure described so far will obtain the optimal ILC amplitudes for the initial computational basis state reference. Simultaneous relaxation of the reference Bloch angles and ILC amplitudes can be accomplished on a classical computer by an multiconfigurational self consistent field (MCSCF)-like two step iterative procedure: 1) solve generalized eigen-problem for subspace Hamiltonian matrix resolved in the $N+1$ states to obtain the current-step optimal ILC amplitudes $\{\tau, \alpha_1, \hdots, \alpha_N \}$, and 2) dress initial Hamiltonian $\hat H$ with the QCC-ILC unitary using obtained amplitudes of the current step
\begin{align}
\tilde{H} = \hat U_{\text{ILC}}^\dagger (\tau, \boldsymbol{\alpha}) \hat H \hat U_{\text{ILC}} (\tau, \boldsymbol{\alpha}),
\end{align} 
then perform Bloch angle optimization with respect to $\tilde{H}$. 
The current-step optimized mean-field state is then used to update the subspace basis of Step 1. The two steps are then repeated until convergence of the energy, yielding the relaxed $N$ ILC amplitudes and $2n_q$ Bloch angles. Over the course of optimization, the reference state may violate $z$-collinearity, resulting in subspace basis states $\ket{\Phi_1} \equiv \ket{\boldsymbol{\Omega}}$, $\ket{\Phi_j} \equiv \hat T_j \ket{\boldsymbol{\Omega}}, j > 1$   no longer being generally orthogonal. If relaxation of QMF is considered, the next iteration of Step $1$ must hence be accomplished by solving non-orthogonal generalization of Eq.~(\ref{eigenproblem}),
\begin{align}
\bar{\boldsymbol{H}} \boldsymbol{c} = E \bar{\boldsymbol{S}} \boldsymbol{c},
\end{align}
where $\bar{\boldsymbol{S}} = \boldsymbol{M} \circ \boldsymbol{S}$ and $\boldsymbol{S}_{ij} = \left \langle \Phi_i | \Phi_j \right \rangle$. 

%


\end{document}